\documentclass[%
 reprint,
superscriptaddress,
amsmath,amssymb,
aps,
]{revtex4-2}

\usepackage{graphicx}
\usepackage{xcolor}
\usepackage{dcolumn}
\usepackage{bm}
\usepackage{comment}
\usepackage{wrapfig}
\usepackage{natbib}

\usepackage{multirow}
\usepackage{makecell}
\usepackage{float}
 
\usepackage{units}
\usepackage{color}
\usepackage{url}

\usepackage[colorlinks]{hyperref}
\hypersetup{%
	plainpages=true,
	breaklinks=true,
	hypertexnames=false,
	pageanchor=true,
	colorlinks=true,
	linkcolor={blue},
	citecolor={red},
	urlcolor={blue},
	anchorcolor={black}
}

\newcommand{\ket}[1]{|#1\rangle}

\begin{document}

\preprint{APS/123-QED}

\title{Engineering symmetry-selective couplings of a superconducting artificial molecule to microwave waveguides} 

\author{Mohammed Ali Aamir}
 \email{aamir.ali@chalmers.se}
\author{Claudia Castillo Moreno}
\author{Simon Sundelin}
\author{Janka Bizn\'arov\'a}
\author{Marco Scigliuzzo}
\author{Kowshik Erappaji Patel}
\author{Amr Osman}
\author{D. P. Lozano}
\author{Simone Gasparinetti}
\email{simoneg@chalmers.se}
 \homepage{https://202q-lab.se}
\affiliation{%
Department of Microtechnology and Nanoscience, Chalmers University of Technology, 412 96 Gothenburg, Sweden
}%

\date{\today}

\begin{abstract}
Tailoring the decay rate of structured quantum emitters into their environment opens new avenues for nonlinear quantum optics, collective phenomena, and quantum communications. Here we demonstrate a novel coupling scheme between an artificial molecule comprising two identical, strongly coupled transmon qubits, and two microwave waveguides. In our scheme, the coupling is engineered so that transitions between states of the same (opposite) symmetry, with respect to the permutation operator, are predominantly coupled to one (the other) waveguide. The symmetry-based coupling selectivity, as quantified by the ratio of the coupling strengths, exceeds a factor of 30 for both the waveguides in our device. In addition, we implement a two-photon Raman process activated by simultaneously driving both waveguides, and show that it can be used to coherently couple states of different symmetry in the single-excitation manifold of the molecule. Using that process, we implement frequency conversion across the waveguides, mediated by the molecule, with efficiency of about 95\%. Finally, we show that this coupling arrangement makes it possible to straightforwardly generate spatially-separated Bell states propagating across the waveguides. We envisage further applications to quantum thermodynamics, microwave photodetection, and photon-photon gates.
\end{abstract}

\maketitle

Waveguide quantum electrodynamics (QED) is an emerging field of research that studies the interaction of quantum emitters with waveguides hosting a continuum of one-dimensional photonic modes~\cite{sheremet2021, roy2017a, lodahl2015, gu2017}. It has been explored in several experimental platforms that include atoms~\cite{corzo2019}, solid-state quantum defects~\cite{sipahigil2016a}, semiconductor quantum dots~\cite{foster2019}, and superconducting circuits~\cite{mirhosseini2019} with either optical or microwave waveguides. A plethora of rich and diverse physics have become accessible from these studies, such as, resonance fluorescence~\cite{astafiev2010}, non-classical states of light~\cite{hoi2012, lu2021c}, collective effects~\cite{devoe1996, guerin2016, solano2017, corzo2019, lalumiere2013, vanloo2013, mirhosseini2019, masson2020, zanner2021}, giant artificial atoms \cite{kockum2018,kannan2020, vadiraj2021}, chiral photonic transport~\cite{lodahl2017, scheucher2016, mahmoodian2020, gheeraert2020a, guimond2020}, atom-photon bound states and interactions with photonic band edges~\cite{liu2016c,mirhosseini2018,sundaresan2019,brehm2021}, topological physics~\cite{kim2021a, besedin2021},  tunable non-Markovian dynamics~\cite{andersson2019a,ferreira2021}. Waveguide QED finds applications in single-photon sources and quantum communication \cite{lodahl2015}, quantum information processing~\cite{kannan2020b}, and recently, quantum thermodynamics~\cite{cottet2017,scigliuzzo2020, lu2021d, monsel2020}.

The primary aspect in waveguide QED is the tailoring of the coupling mechanisms of quantum emitters to a waveguide.
When multiple, resonant emitters are involved, they form collective states, known as Dicke states~\cite{dicke1954b}. The emission properties of Dicke states
are determined by the symmetry properties of their composite wavefunction, with respect to exchanges of emitters.
Depending on the symmetry, Dicke states thus appear either super-radiant or sub-radiant, i.e., they emit rapidly or slowly, 
and are
referred to as bright and dark states, respectively~\cite{devoe1996, gonzalez-tudela2013, vanloo2013, mlynek2014a}. 
Owing to their isolated nature, dark states are promising resources for quantum information processing~\cite{monz2009a, kielpinski2001, paulisch2015, lidar1998} and quantum memories~\cite{leung2012}. However, for the same underlying reason, they are also challenging to control or detect~\cite{guerin2016, solano2017, zanner2021}. Quantum control of dark states has been achieved using multiple classical drives with a definite phase relation
~\cite{filipp2011, zanner2021}; however, due to their dark nature, their excitation could not be measured directly.


\begin{figure*}
\includegraphics[width=1\linewidth]{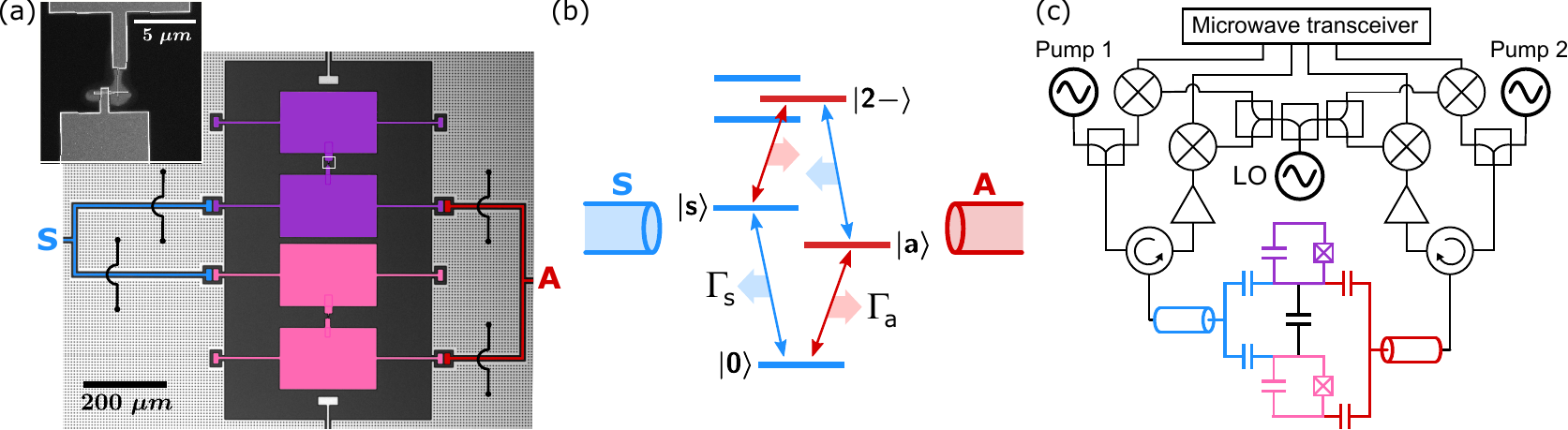}
\caption{Device architecture and experimental set-up. (a)~False-color micrograph of the device comprising two split transmons (pink and violet) coupled strongly to two microwave waveguides labelled~A (red) and~S (blue). Black wires indicate the positions of bonding wires used to connect isolated islands to the ground plane. A small white box delineates the area containing one of the Josephson junctions which has been enlarged in the inset. Additional (uncolored) coupling conductors at the top and bottom of the micrograph are part of resonators not used in the experiments. (b)~Energy-level diagram of the collective states of the two-transmon system up to the double-excitation manifold.
States with even (odd) symmetry, as well as symmetry-preserving (symmetry-inverting) transitions are indicated in blue (red). symmetry-preserving (symmetry-inverting) transitions predominantly couple to waveguide~S~(A), as indicated by horizontal arrows. The transitions to the other doubly-excited states are omitted for clarity [see SI for more details].
(c)~Simplified experimental setup (LO: local oscillator). See text and SI for details.
}
\end{figure*}

Here, we present a unique yet simple architecture of a superconducting artificial molecule coupled to two microwave waveguides such that each waveguide couples \textit{selectively} to one of the two manifesting symmetries, characterized by the permutation operator, of the collective states of the molecule. Thus, each collective state with of each symmetry is a dark state to one waveguide but a bright state to the other waveguide, and therefore is amenable to independent detection. In addition, our scheme has a lower hardware requirement for quantum control of the dark state as it does not require any static or dynamic control of the applied phase as in the previous works~\cite{filipp2011, zanner2021}. In this work, we demonstrate two distinct experiments. In the first one, we couple the bright and dark state by activating an efficient two-photon Raman process via a doubly-excited state and thereby mediate coherent population transfers between the two states with opposite symmetries at an efficiency of about 95\%~\cite{gerry1990}; thereby, also achieving a frequency conversion~\cite{roch2012}. In the second experiment, we generate spatially separated, entangled itinerant photons, in particular a Bell state, using a remarkably simple scheme. This capability~\cite{menzel2012, flurin2012, gasparinetti2017, peugeot2021} is of significance for the application of waveguide QED for distributing quantum entanglement, especially for the purpose of distributed quantum information processing at spatially separated quantum processor nodes~\cite{narla2016, kurpiers2018, axline2018, campagne-ibarcq2018, kannan2020b}.

The device contains an artificial molecule made of two nominally identical, mutually-interacting artificial atoms, each realized with a superconducting transmon \cite{koch2007a}. Each transmon has two pads forming its capacitor in parallel to a Josephson junction [Fig.~1(a)]. They are coupled to two waveguides in a novel geometry, such that, waveguide named~S~(A) is coupled to the inner (same-side) pads of each transmon. As a result, the drive operators caused by waveguides~S and~A in the rotating frame are directly proportional to $\hat b_1 + \hat b_2 + \hat b^\dagger_1 + \hat b^\dagger_2 $ and $\hat b_1 - \hat b_2 + \hat b^\dagger_1 - \hat b^\dagger_2$, respectively, where $\hat b_i$ is the annihilation operator of transmon $i=\{1,2\}$. The Hamiltonian governing the molecule is $\hat H=\sum_{i=1,2} (\omega_i \hat b^\dagger_i \hat b_i + \alpha_i \hat b^\dagger_i \hat b^\dagger_i \hat b_i \hat b_i/2) + g(\hat b^\dagger_1 \hat b_2 + \hat b_1 \hat b^\dagger_2)$, where $\omega_i/2\pi$ and $\alpha_i$ are the mode transition frequency and the anharmonicity of transmon $i$; $g$ is the inter-transmon coupling rate~\cite{koch2007a}. The Hamiltonian is invariant under the exchange of its constituent transmons, therefore, it commutes with the permutation operator~\cite{filipp2011a, begzjav2019} whose eigenvalues are $\pm 1$. Thus, the eigenstates of the Hamiltonian yield an eigenvalue of either +1 or -1 with the permutation operator, are thus either symmetrical or anti-symmetrical, respectively. The molecule's bare modes in the single-excitation manifold, the states $|01\rangle$ and $|10\rangle$, are resonant and so split by $2g$ into collective states that are orthogonal linear combinations $|s\rangle = (|01\rangle + |10\rangle)/\sqrt{2}$ and $|a\rangle = (|01\rangle - |10\rangle)/\sqrt{2}$ that are symmetrical and anti-symmetrical, respectively [Fig.~1(b)]. The double-excitation manifold has three collective states where one of them, $|2-\rangle = (|02\rangle - |20\rangle)/\sqrt{2}$, is anti-symmetrical and the other two are symmetrical [see SI]. Owing to the form of their drive operators, the waveguides~S~(A) cause dipole moment transitions between the collective states that is symmetry-preserving (symmetry-inverting), dictating symmetry-based selection rules [see transition arrows in Fig.~1(b)]. For instance, the transitions $\ket{0} \leftrightarrow \ket{s}$ ($\ket{0} \leftrightarrow \ket{a}$) couples strongly to waveguide~S~(A), at a coupling rate $\Gamma_s$ ($\Gamma_a$). 

We perform all our experiments when the device is at about 9~mK in a dilution refrigerator. Each waveguide is connected to an input line used to deliver microwave drives and an output line with linear amplifiers to boost the scattered microwave signals, separated by a microwave circulator [see Fig.~1(c)]. The input signals consists of a continuous microwave tone, a time-resolved probe tone or a combination of both. The continuous tone is generated by microwave sources, called pump in Fig.~1(c), whereas the time-resolved probe tones are tailored with a microwave transceiver used in conjunction with an up-conversion in-phase-quadrature (IQ) mixer. The microwave transceiver is essentially made of arbitrary waveform generators and digitizers synchronized internally. The digitizer acquires the time-resolved scattered signals as time traces. To perform coherent measurements, especially of population transfers (discussed below) between states $|a\rangle$ and $|s\rangle$ which have different frequencies, it is vital to acquire the time traces with the same phase across multiple realizations (shots) of the experiment. This is enforced by performing all the up/down-conversion with the same local oscillator. More details of the set-up are present in SI.

\begin{figure}
\includegraphics[width=1\linewidth]{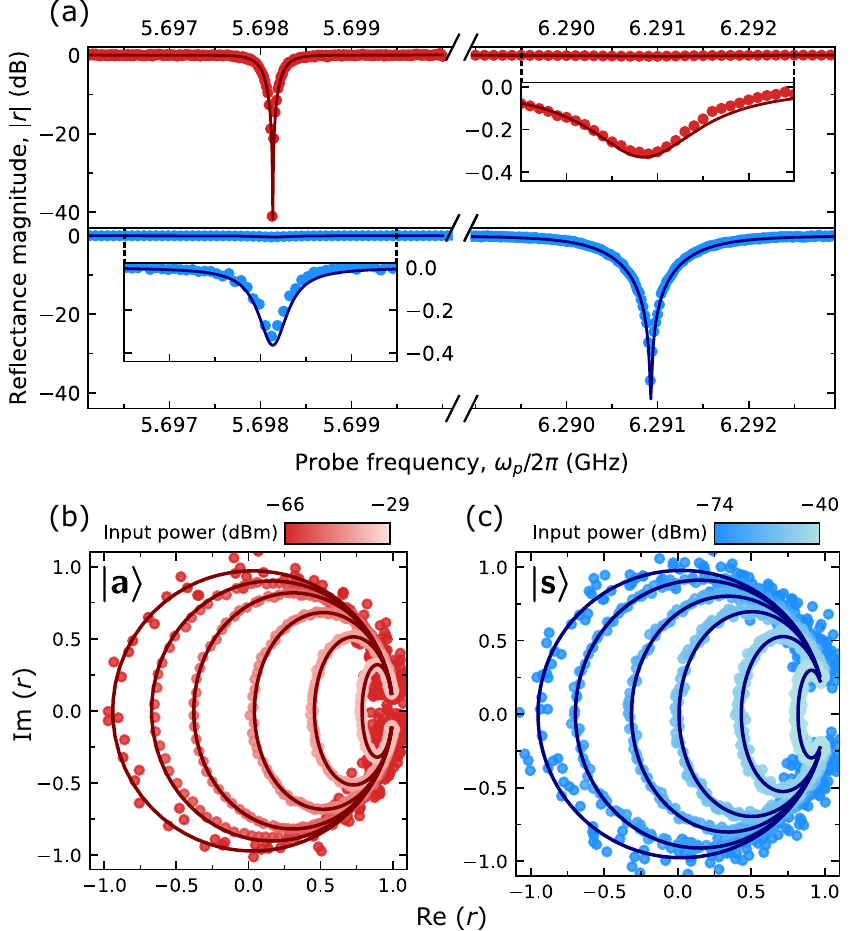}
\caption{Single-tone reflection spectroscopy. (a)~Reflectance magnitude $|r|$ as a function of probe frequency shows signature of both the state $|a\rangle$ at $\omega_a/2\pi = 5.6981$~GHz and state $|s\rangle$ at $\omega_s/2\pi=6.2909$~GHz when probed from the waveguides A (in upper panel) and S (in lower panel). Plot of imaginary part versus real part of the reflectance for (b)~the state $|a\rangle$ probed in waveguide~A and (c)~state $|s\rangle$ in probed in waveguide~S at a few input powers. All solid lines are fits based on the model in the text.}
\end{figure}

We first measure the power-dependent reflectance $r$ from both waveguides using a vector network analyzer (Fig.~2). It is strongly dependent on the coupling rate between each waveguide and the single-excitation state of corresponding symmetry, $\Gamma_{\{s,a\}}$, and the coupling rate of that state to any other decay channel, $\Gamma'_{\{s,a\}}$. When measured from waveguide~A, $|r|$ exhibits a strong suppression at the mode frequency of $\ket{a}$, $\omega_a/2\pi$, at a power referred to as the ``magic power'' \cite{hoi2015}, a signature that the transition $\ket{0} \leftrightarrow \ket{a}$ is over-coupled to waveguide A ($\Gamma_a > \Gamma'_a$). The suppression is due to destructive interference between reflected and coherently scattered radiation from a two-level system. A small dip ($\sim 0.3$ dB) also appears at the mode frequency of $\ket{s}$, $\omega_s/2\pi$, at sufficiently low powers, indicating that the transition $\ket{0} \leftrightarrow \ket{s}$ is strongly under-coupled. When measuring $r$ from waveguide~S, we make opposite observations, i.e.,  we observe full suppression at $\omega_s/2\pi$ but only a small dip at $\omega_a/2\pi$ [Fig.~2(a), bottom]. For the two over-coupled transitions, we measure the full power dependence of $r$ and find that at low power $r$ describes a nearly unit circle in the in-phase-quadrature (IQ) plane, which continuously reduces towards a single point (+1) as the power is increased [Figs.~2(b,c)]. This trend reflects the transition between coherent and incoherent scattering as a two-level system is driven towards saturation, as observed before \cite{scigliuzzo2020}. The data is very well described by a two-level model based on Linblad master equation and input-output theory \cite{lalumiere2013} (see SI). Based on global fits of the model to the data taken at each resonance~\cite{lu2021b, hoi2015}, we extract the coupling rates $\Gamma_{\{s,a\}}$ and $\Gamma'_{\{s,a\}}$ (Table~1). We find that $\Gamma'_{\{s,a\}}$ is largely the coupling to the waveguide of opposite symmetry because the corresponding $r$ measurements can be fit fairly well by exchanging the values of $\Gamma_{\{s,a\}}$ and $\Gamma'_{\{s,a\}}$ [Fig.~2(a), both insets; also see SI]. State $\ket{s}$ ($\ket{a}$) primarily emits into waveguide~S~(A) with selectivity $\Gamma_s/\Gamma_s' = 47$ ($\Gamma_a/\Gamma_a' = 35$), demonstrating a very high symmetry selectivity of the dipole moment transitions mediated by the waveguides.\\

\begin{table}
\centering
    \begin{tabular}{ l c c }
    \hline
    Parameter & Symbol & Value \\
    \hline
    $\ket{s}$ mode frequency & $\omega_a/2\pi$ & 5.6981 GHz\\ 
    $\ket{a}$ mode frequency & $\omega_s/2\pi$ & 6.2909 GHz\\
    Qubit-qubit coupling  & $g/2\pi$ & 296.4 MHz\\
    $\ket{s}\to\rm{S}$ decay rate & $\Gamma_s/2\pi$ & 1.388 MHz\\
    $\ket{a}\to\rm{A}$ decay rate &$\Gamma_a/2\pi$ & 0.311 MHz\\
    $\ket{s}\to\rm{A}$ decay rate &$\Gamma_s'/2\pi$ & 29.8 kHz\\
    $\ket{a}\to\rm{S}$ decay rate &$\Gamma_a'/2\pi$ & 8.8 kHz\\
    \hline
    \end{tabular}
    \caption{Measured parameter values. See text for details.}
    \end{table}

\begin{figure}
\includegraphics[width=1\linewidth]{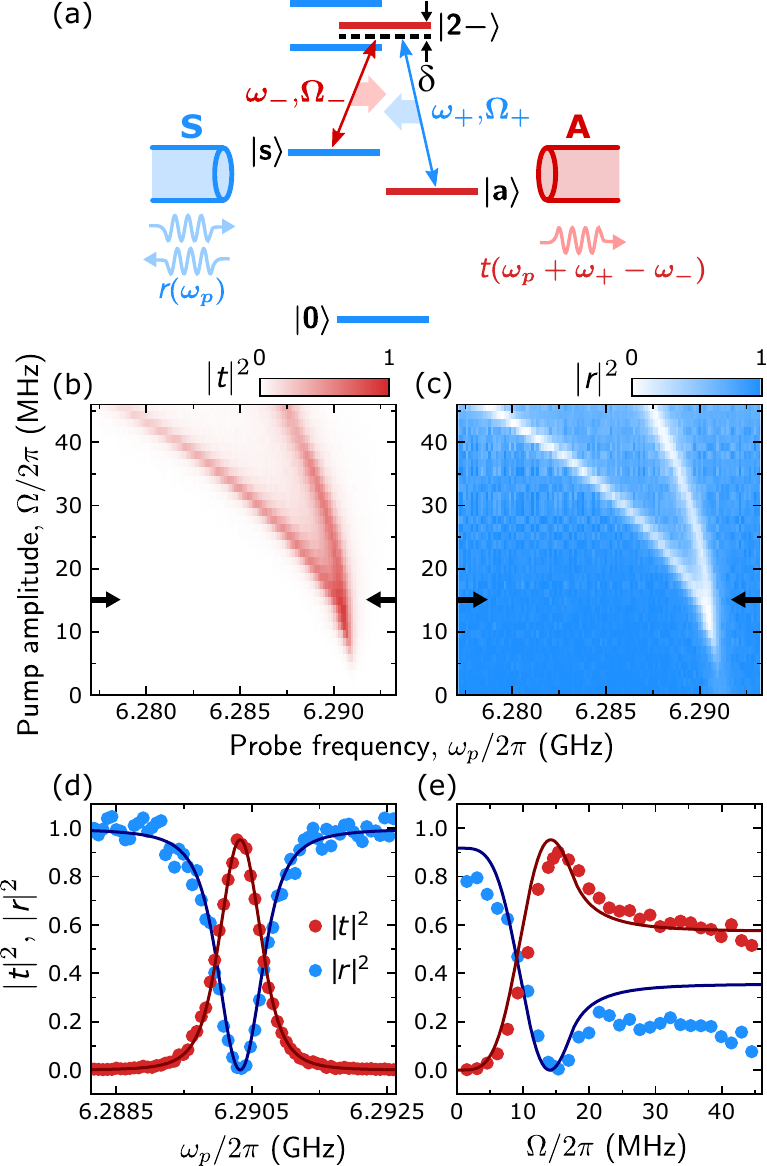}
\caption{Frequency converter based on coherent Raman process.
(a) Level diagram.
A Raman process is enabled by two microwave pump tones driving the transitions between state $|a\rangle$ ($|s\rangle$) and a virtual state detuned from state $|2-\rangle$ by $\delta$ at frequency $\omega_+$ ($\omega_-$) and pump amplitude $\Omega_+$ ($\Omega_-$) mediated by the waveguide~S~(A). The resulting coherent population transfer from state $|s\rangle$ to state $|a\rangle$ is observed by probing the reflectance, $r$, while sweeping the probe frequency $\omega_p$ of a very weak probe tone sent to waveguide~S and simultaneously measuring the coherent transmittance, $t$, in waveguide~A at the locked frequency $\omega_p+\omega_+-\omega_-$.
(b,c) 2D plot of (b)~power transmittance, $|t|^2$, and (c)~power reflectance, $|r|^2$,  as a function of $\omega_p$ and pump amplitude $\Omega=\Omega_+=\Omega_-$.
(d)~Plot of $|t|^2$ and $|r|^2$ at the optimal $\Omega = 15.35$~MHz as a function of $\omega_p$. (e)~Plot of the maximum $|t|^2$ and correspondingly minimum $|r|^2$ as a function of $\Omega$, when following the right branch in (c). Solid lines in (d,e) are theory predictions (see text for details).}
\end{figure}

The selectivity in the emission properties of states with opposite symmetry is desirably complemented by a mechanism to activate a strong, coherent coupling between them. This is enabled by
a two-photon Raman process between $\ket{s}$ and $\ket{a}$, mediated by the state $|2-\rangle$~\cite{gerry1990, bergmann1998a, kumar2016a} [Fig.~3(a)].
To do so, we send pump tones at frequency $\omega_+$ ($\omega_-$) with pump amplitude $\Omega_+$ ($\Omega_-$) to waveguide~S~(A). The Raman resonance occurs when $\omega_+ - \omega_- \approx \omega_s - \omega_a$ and, in the doubly-rotated frame, causes the orthogonal states $|a\rangle$ and $|s\rangle$ to become resonant and couple with strength $\Omega_+ \Omega_- / 2\delta$~\cite{gerry1990}, where $\delta$ is the detuning between the virtual state of the Raman process and $\ket{2-}$. The pumping process is highly efficient due to the large dipole moments for the transition $|a\rangle \leftrightarrow |2-\rangle$ from waveguide~S and transition $|s\rangle \leftrightarrow |2-\rangle$ from waveguide~A as favoured by the selection rules.
To observe the coupling, we study the transmission of weak coherent tones of frequency $\omega_p$ and drive amplitude $\Omega_p$ from waveguide S into waveguide A, at the converted frequency $\omega_t = \omega_p + \omega_+ - \omega_-$, mediated by states $\ket{s}$ and $\ket{a}$ and by the Raman coupling.
The phase coherence has been enforced across all channels in the measurement setup by using the same local oscillator to drive all four up/downconversion mixing stages~[Fig.~1(c)]. We fix $\delta/2\pi = 300$~MHz, a sufficient detuning to avoid direct population of the $\ket{2-}$ state.
The measured power transmission, $|t|^2$, as a function of $\omega_t$ exhibits a clear peak, signifying Raman resonance, when the common pump amplitude $\Omega=\Omega_{+,-}$ is sufficiently increased and reaches nearly 90.1\% power transmission at $\Omega/2\pi = 15.35$~MHz~[Fig.~3(b)]. Increasing beyond this optimal value, $\Omega$ becomes large enough to split the resonant Raman levels that appear as two local maxima in $|t|^2$, while Stark-shifting the resonance at the same time. The same trends are observed in the measured power reflection, $|r|^2$, which exhibits the corresponding minima where the Raman resonance occurs~[Fig.~3(c)]. The efficiency of the population transfer is sensitive to the probe power used, which is about $\Omega_p/\Gamma_s = 0.116$ for the measurement of Fig.~3(b-c). When using a smaller probe amplitude, $\Omega_p/\Gamma_s = 0.077$, well in the linear response regime
the maximum $|t|^2$ reaches 95.2\%~[Fig.~3(d)]. $|r|^2$ diminishes to almost zero here, implying that less than 5\% of power is lost incoherently during optimal population transfer. Beyond the optimal $\Omega$, the maximum achievable $|t|^2$ decreases slowly with $\Omega$ [Fig.~3(e)].
In Fig.~3(e), we note that $|r|^2$ is lower than 1 for $\Omega \to 0$ because of both finite direct scattering into the waveguide~A ($\Gamma_s'/\Gamma_s>0$) and partial saturation due to a non-zero probe amplitude $\Omega_p$. These results are in excellent agreement with a simple two-state model based on a non-Hermitian Hamiltonian [solid lines in Fig.~3(e); see SI for details].  
The model uses only independently extracted spectroscopic parameters, without any fitting parameters.

The $V$-shaped structure of the level diagram in Fig.~1(b) lends itself well to generating entanglement between the radiation emitted by the $\ket{s}\to{\ket{0}}$ and $\ket{a}\to{\ket{0}}$ transitions. Thanks to our waveguide engineering technique, this radiation is directly emitted into spatially separated modes.
To demonstrate entanglement between propagating photonic modes,
We enable the following sequence of events
[Fig.~4(a)]:
(1) 
induce a $\pi/2$ rotation between $\ket{0}$ and $\ket{a}$ by applying a resonant pulse to waveguide~A;
(2) induce a rotation of variable angle $\theta$ between $\ket{0}$ and $\ket{s}$ by applying a resonant pulse to waveguide~S; and
(3) let the molecule spontaneously decay, thereby transferring the original entanglement to propagating photonic modes.
After the above sequence, we expect to find the system in the state
\[ \frac{1}{\sqrt{2}}|0\rangle\left[\cos \frac{\theta}{2}|0\rangle_{A}|0\rangle_{S}+\left( \sin \frac{\theta}{2}|0\rangle_{A}|1\rangle_{S}+|1\rangle_{A}|0\rangle_{S}\right)\right]\]
where $\ket{\{0,1\}}_{\{S,A\}}$ are Fock states in propagating modes of waveguides S and A, which in the following we describe by the annihilation operators $\hat a_+$ and $\hat a_-$, respectively. 

We perform tomographic reconstruction of selected moments of the propagating photonic modes, $\hat a_{+,-}$, which are simultaneously read-out on the output lines of both waveguides using our linear amplification chain and temporal mode matching, employing the techniques described in \cite{eichler2011a} (see SI for more details). We also compare our results against a control experiment in which event (1) is omitted from the sequence.

The measured first-order moments $\langle \hat{a}_- \rangle$ and $\langle \hat{a}_+ \rangle$ [circle data points in Fig.~4(b) and~4(c)] are close to the expectation values, $\langle \hat{a}_- \rangle = \frac{1}{2} \cos{\theta/2}$ and $\langle \hat{a}_+ \rangle = \frac{1}{4} \sin{\theta}$ calculated on the full quantum state above [solid lines in Fig.~4(b),~4(c)]. However, the measurements deviate slightly from these ideal functional forms because of the short lifetimes of these states ($1/\Gamma_a = 512$~ns, $1/\Gamma_s = 115$~ns) and very small cross-coupling of the states to the waveguide of the opposite symmetry. These results are compared against the case when no $\pi/2$ pulse is sent to state $|a\rangle$ [square data points in Fig.~4(b) and~4(c)]. We expect the photon flux $\langle \hat{a}_-^\dagger \hat{a}_- \rangle$ to be 0.5, irrespective of $\theta$ [Fig.~4(d)] because the $\pi/2$ pulse in waveguide~A drives the molecule to the superposition state $(|0\rangle + |a\rangle)/\sqrt{2}$. This data has been used to determine the normalization coefficient for the photonic mode $\hat{a}_-$. Following the $\pi/2$ pulse, the remaining population of molecule's state $|0\rangle$ is nearly 0.5 which is available for coherent exchange with the state $|s\rangle$ depending on $\theta$ of the second pulse. Consequently, $\langle \hat{a}_+^\dagger \hat{a}_+ \rangle$ oscillates upto an amplitude of nearly 0.5 as opposed to 1 when $\pi/2$ pulse is absent and state $|0\rangle$ population is 1 to begin with [Fig.~4(e)]. This data was used to normalize the photonic mode $\hat{a}_+$. All normalizations have taken into account the decay of the population prior to the read-out due to short lifetime of the states.

\begin{figure}
\includegraphics[width=1\linewidth]{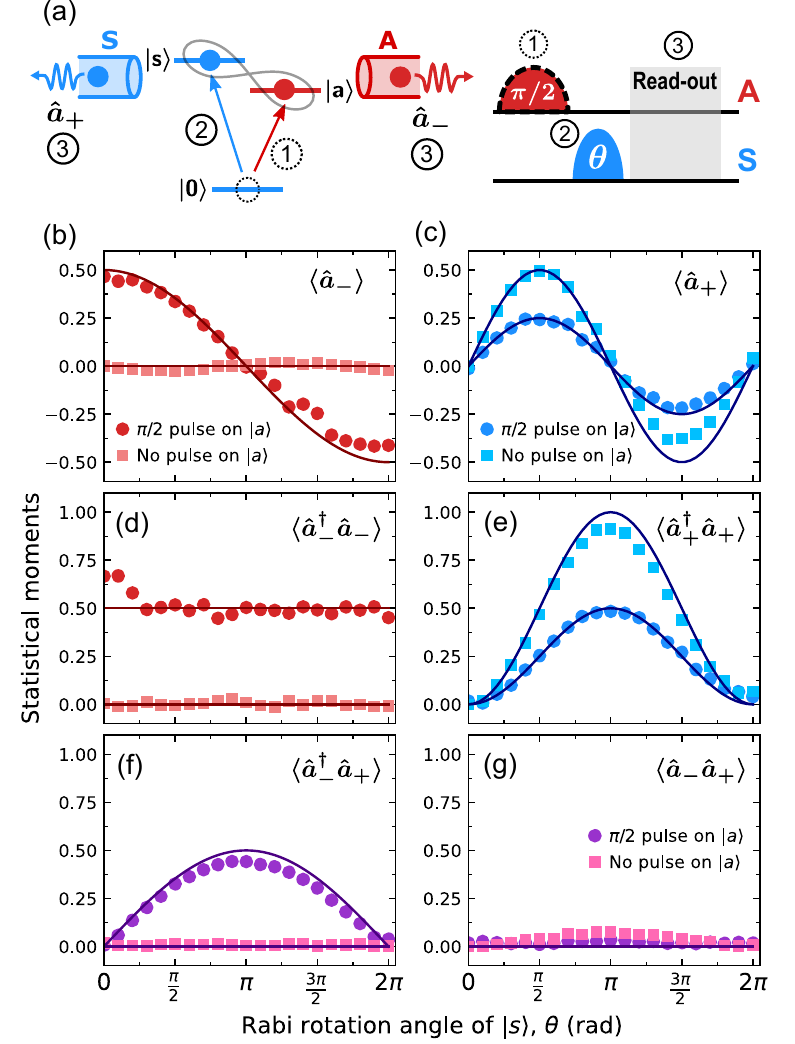}
\caption{Entanglement of propagating microwave fields in two separate waveguides. (a) (Left) Step sequence to entangle propagating fields in the waveguides~A and~S where the labels are same as the description in the text. (Right) Corresponding pulse sequence followed by simultaneous read-outs in the output lines of both waveguides. The entanglement is conditional on the first step where $\pi/2$ pulse is sent to state $|a\rangle$, so all results are compared against the case with an absent $\pi/2$ pulse in waveguide~A.
(b-g) Statistical moments of the propagating modes $\hat{a}_-$ in waveguide~A and $\hat{a}_+$ in waveguide~S
\textit{vs}
Rabi rotation angle $\theta$ of the pulse sent to waveguide~S.
}
\end{figure}

The main signature of entanglement is encoded in the cross-moments, $\langle \hat{a}_-^\dagger \hat{a}_+ \rangle$ and $\langle \hat{a}_- \hat{a}_+ \rangle$
[Fig.~4(f) and~4(g)]. Only $\langle \hat{a}_-^\dagger \hat{a}_+ \rangle$ takes a sufficiently large value when the $\pi/2$-pulse is applied to $|a\rangle$,  and reaches nearly 0.5 at $\theta = \pi$ following the functional form $\langle \hat{a}_-^\dagger \hat{a}_+ \rangle = \frac{1}{2} \sin{\theta/2}$ as expected for the Bell state $(|01\rangle + |10\rangle)/\sqrt{2}$ from the full quantum state above. In addition, a much smaller value of the other combination of the moments, $\langle \hat{a}_- \hat{a}_+ \rangle$ gives additional assurance of the formation of this specific Bell state. The slight variation close to zero is likely due to the small leakages we noted in the other moments above. While full quantum state tomography \cite{eichler2011a} would be required in order to quantify the fidelity of the generated states, the measured data provide compelling evidence that the targeted process is realized in our system, also given that resonant driving of a two-level system followed by spontaneous decay has been used in previous work to implement highly efficient single-photon sources \cite{peng2016,pechal2016a}.

In summary, we have presented a novel waveguide QED architecture in which transitions in a diatomic artificial molecule are selectively coupled to two waveguides, depending on their inherent symmetries.
The selective coupling mechanism is implanted in the device and requires no static or dynamic tuning of frequencies or phase differences in control pulses \cite{filipp2011, zanner2021}.
We have provided two examples of the capabilities of this architecture:
a coherent frequency-converter operating across the two waveguides with efficiency close to unity, and a simple scheme to generate maximally entangled propagating modes in spatially separate waveguides.
A number of further applications can be envisaged.
Setting one of the two waveguides in the undercoupling regime enables the creation of a long-lived metastable state. The metastable state can be coupled to the bright state either coherently, via a Raman process, or irreversibly, exploiting resonant coupling to the second-excitation manifold followed by photon emission into the strongly coupled waveguide. The latter scheme leads to an impedance-matched lambda system that can be exploited for photodetection \cite{inomata2016,lescanne2020a}.
A scheme similar to the one used to generate Bell states may be used to realize photon-photon gates \cite{reuer2022} across separate waveguides.
When the waveguides are populated with thermal fields \cite{scigliuzzo2020,lu2021d}, and the dark and bright states are coherently coupled by demonstrated Raman process, the molecule operates as a quantum thermal machine (heat engine or refrigerator), paving the way for studies in quantum thermodynamics. Finally, the presented scheme can be extended to larger artificial molecules, arrays of qubits or resonators \cite{yanay2021}, and generally to any other physical system that enables near-field coupling to both ends of an electric or magnetic dipole, for example, laterally defined double quantum dots \cite{stockklauser2017}.

We are grateful to Francesco Ciccarello, Giuseppe Calajò and Anton Frisk Kockum for useful feedback to our results. The presented device was fabricated in Myfab Chalmers, a nanofabrication laboratory, and its design was assisted by the python package QuCat~\cite{Gely2020}. We thank the Swedish Research Council, the Knut and Alice Wallenberg Foundation through the Wallenberg Center for Quantum Technology (WACQT) for financial support.

See Supplemental Material for (1) full experimental set-up (2) single tone reflection spectroscopy's modelling and supplementary data (3) pump amplitude calibration (4) details of tomographic reconstructions of selected field moments (5) full Hamiltonian and the transition dipole matrix from both waveguides (6) Basic two-photon Raman model.


%


\clearpage
\onecolumngrid
\appendix

\renewcommand\thefigure{S\arabic{figure}}    
\renewcommand\thetable{S\arabic{table}}
\renewcommand\theequation{S\arabic{equation}}    
\setcounter{figure}{0} 
\setcounter{equation}{0}
\setcounter{table}{0}

\section*{Supplemental material: Engineering symmetry-selective couplings of a superconducting artificial molecule to microwave waveguides} 

\subsection{Full experimental setup}

\begin{figure*}[h]
\includegraphics[width=1\linewidth]{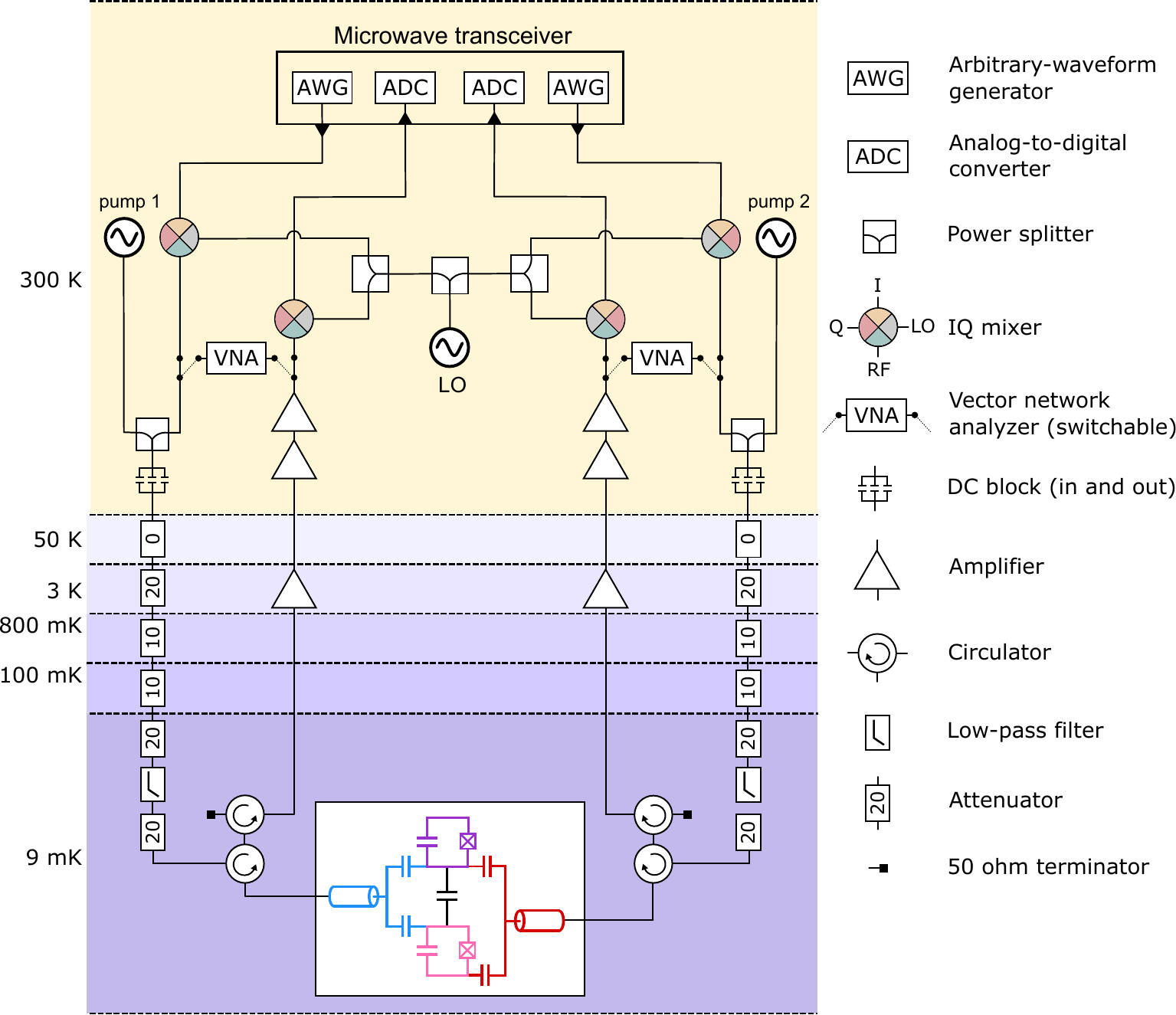}
\caption{Full experimental set-up. See text for description. The in-phase-quadrature (IQ) mixers'  ports~I and Q are both used to connect to the microwave transceiver, but only port~I is shown to be wired while omitting port~Q for clarity.
}
\end{figure*}

Fig.~S1 shows the full experimental setup used in the experiment. The device is thermally anchored at the mixing chamber stage of a dilution refrigerator that reaches 9~mK. The device is enclosed in a copper box, and further shielded against electromagnetic waves by a copper enclosure, and against low-frequency magnetic field by a $\mu$-metal enclosure. Classical fields are sent through highly attenuated input coaxial lines, whereas the signals from the device are collected in output lines which is equipped with a cryogenic HEMT amplifier (provided by Low Noise factory) at 3~K. The signal is further boosted by room temperature ($\approx$~300~K) amplifiers. The input and output signals are routed by microwave circulator. The measurements are performed either by either a vector network analyzer (VNA) or a microwave transceiver in conjunction with an in-phase-quadrature (IQ) mixers, that have been physically toggled between (represented by a switch schematic in Fig.~S1). The VNA is used for continuous-wave reflection spectroscopy that relies on the scattering properties of the artificial atoms. It has been used in single-tone spectroscopy to find the mode frequency of the states $\ket{a}$ and $\ket{s}$, and also for two-tone spectroscopy to determine the transitions to higher levels when used in combination with a pump tone provided by an microwave source. The pump tone is superimposed using a power splitter (that also works as a power combiner).

The microwave transceiver (Vivace board from Intermodulation Product) is used for time-resolved measurements. It is composed of arbitrary waveform generators (AWG) and analog-to-digital converters (ADC, also called digitizers) that operate in sync with respect to both timing and phase. Furthermore, it is also augmented by FPGA logic which enables interleaved measurements, among other advantages. The AWG produces the pulses of arbitrary envelope shapes at an intermediate carrier frequency near 300~MHz, which is up-converted by the IQ mixer before delivered to the device. The output signals received from the device are amplified, down-converted back to the same intermediate frequency and then finally read-out by the ADC. Very importantly, all the up-conversion/down-conversion are driven by the same local oscillator (LO) operating near 6~GHz, so that they have the same phase on each repetition (shot) of the measurement before being averaged. This is particularly critical for transmission measurements when operating as a frequency converter.

\subsection{Single-tone reflection spectroscopy: model and supplementary data}

\begin{figure*}
\includegraphics[width=0.5\linewidth]{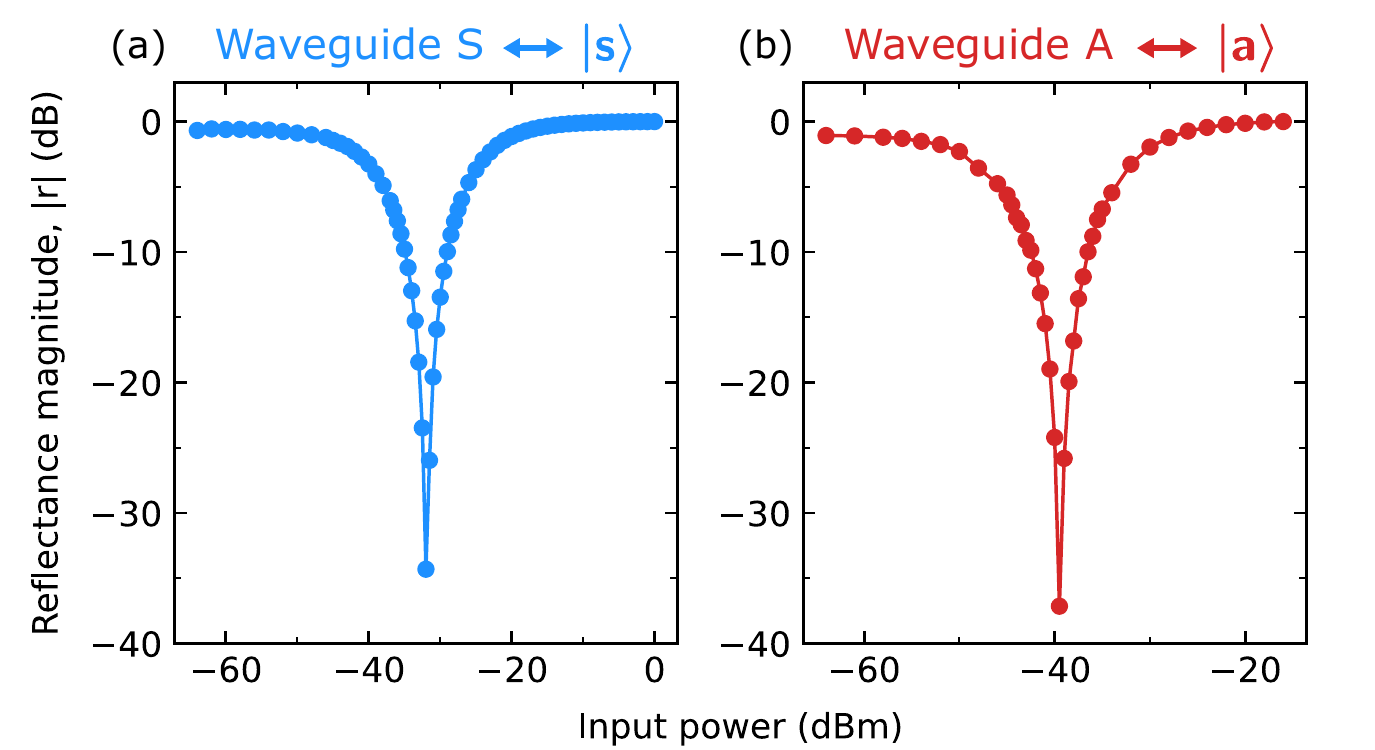}
\caption{Input power dependence of reflectance at zero detuning $\delta = 0$~GHz for (a) $\ket{s}$ and (b) $\ket{a}$, measured from their corresponding over-coupled waveguide.
}
\end{figure*}


Reflectance, $r$, as obtained from scattering from each two-level system at the end of a waveguide can be theoretically calculated using Linblad-based Master equation and input-output theory~\cite{hoi2015, lu2021b, scigliuzzo2020}. In the case of our diatomic molecule, for each state $i = \{s,a\}$, we model the data with

\begin{equation}
r(\omega_p-\omega_i)=1-\frac{i \Gamma_{i} \Gamma_{1i}\left(\omega-\omega_i-i \Gamma_{2i}\right)}{\Omega_i^{2} \Gamma_{2i}+\Gamma_{1i}\left[(\omega-\omega_i)^{2}+\Gamma_{2i}^{2}\right]}
\end{equation}

Here, $\omega_p$ is the probe frequency, $\omega_i$ is the mode frequency, $\Gamma_i$ is the coupling rate of the state $i$ to the probing waveguide and $\Gamma'_i$ is effective coupling rate of state $i$ to all channels other than the probing waveguide; $\Gamma_{1i} = \Gamma_{i}+\Gamma'_{i}$; $\Gamma_{2i} = (\Gamma_{i}+\Gamma'_{i})/2 + \Gamma_{i\phi}$ where $\Gamma_{i\phi}$ is the pure dephasing rate.

Fig.~S2 show the plots of $|r|$ as a function of input power at resonant drive $\omega_p = \omega_{s,a}$, from their respective over-coupled waveguide. The strong suppression of $|r|$ occurs at the ``magic power'' as discussed in the main text.


\subsection{Pump amplitude calibration}

\begin{figure*}[h]
\includegraphics[width=1\linewidth]{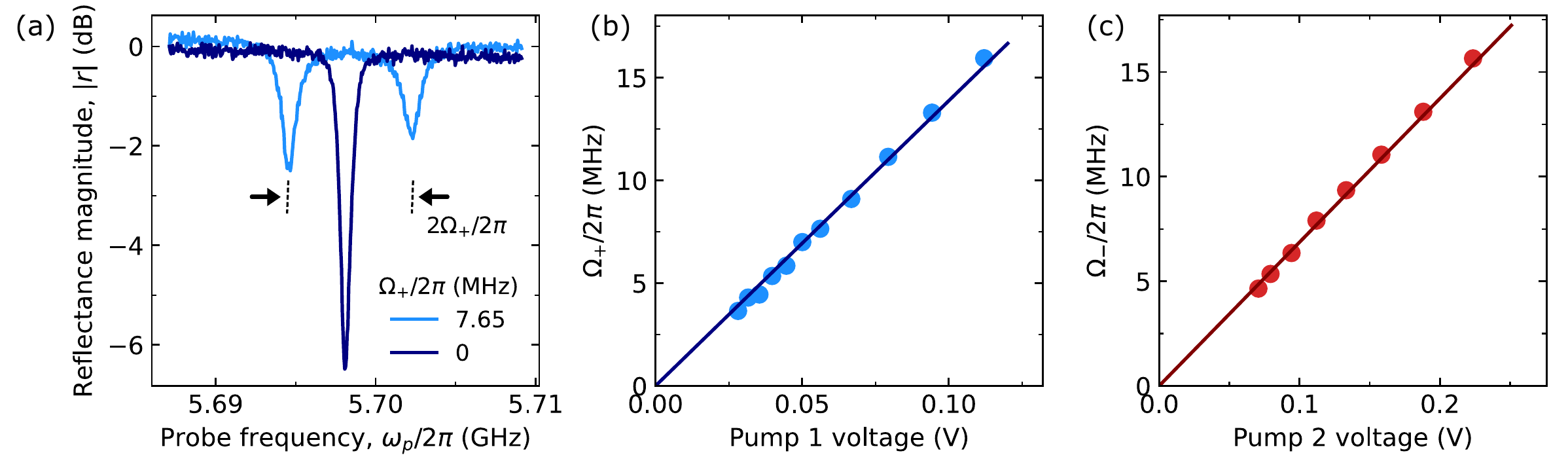}
\caption{Pump amplitude calibration using Autler-Townes splitting. (a) Reflectance magnitude as a function of probe frequency measured on waveguide~A while pumping the transition $\ket{a} \leftrightarrow \ket{2-}$ from waveguide~S with pump 1 at two different pump amplitudes $\Omega_+$. (b, c) The observed splitting (b) $\Omega_{+}/2\pi$ (c) $\Omega_{-}/2\pi$ as a function of the voltage amplitude of the continuous wave microwave tone from their respective pumps. Solid lines in (b,c) are fits.
}
\end{figure*}

The pump amplitudes $\Omega_{+,-}$ reaching the waveguides are calibrated using the phenomenon of Autler-Townes splitting. To calibrate $\Omega_{+}$ ($\Omega_{-}$) sent to waveguide~S~(A), it is used to drive the transitions $\ket{a} \leftrightarrow \ket{2-}$ ($\ket{s} \leftrightarrow \ket{2-}$) while probing the transition $\ket{s} \leftrightarrow \ket{0}$ ($\ket{a} \leftrightarrow \ket{0}$) with waveguide~A~(S) using the vector network analyzer. The splitting increases linearly with the square root of pump power, equivalently, the voltage amplitude of the microwave tone.  Fig.~S3(a) shows the splitting corresponding to $\Omega_{+}/2\pi=7.65$~MHz for waveguide~S. Fig.~S3(b,c) show the observed splitting with increasing pump tone voltage amplitude and corresponding fits which aid in calibrating the pump amplitudes.

\subsection{Propagating field moments: measurement technique and analysis}

\begin{figure*}[h]
\includegraphics[width=0.6\linewidth]{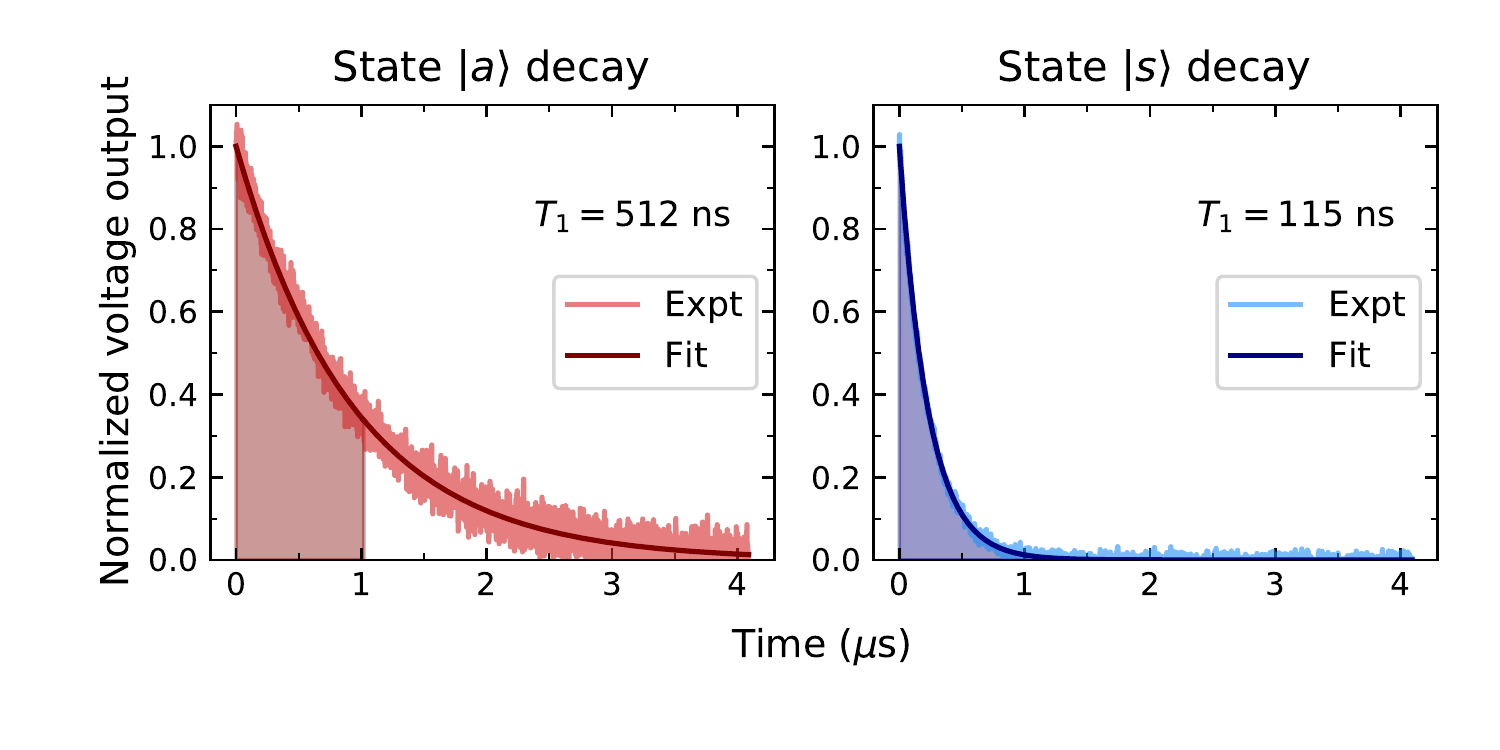}
\caption{Temporal profile of the photonic modes radiated into the waveguide. (Left) state $\ket{a}$ decay into waveguide~A; (right) state $\ket{s}$ decay into waveguide~S. The fit is an exponential decay function with the rate $2T_1$ where $T_1 = 1/\Gamma_{s, a}$ is the energy relaxation time. The shaded area indicates the acquisition duration of $1.02~\mu$s allowed by our hardware over which the temporal mode matching was performed.
}
\end{figure*}


During the step sequence used for generating entangled photonic modes propagating in the two waveguides, the quantum state of the full system goes through the following transformation,

\begin{eqnarray*}
&&|0\rangle\otimes|0\rangle_{A}\otimes|0\rangle_{S} \xrightarrow{\text{(1)}} \frac{1}{\sqrt{2}}\left(|0\rangle+|a\rangle\right)\otimes|0\rangle_{A}\otimes|0\rangle_{S} \xrightarrow{\text{(2)}} \frac{1}{\sqrt{2}}\left(\cos \frac{\theta}{2}|0\rangle+\sin \frac{\theta}{2}|s\rangle+|a\rangle\right)\otimes|0\rangle_{A}\otimes|0\rangle_{S} \\
&& \xrightarrow{\text{(3)}} \frac{1}{\sqrt{2}}(\cos \frac{\theta}{2}|0\rangle\otimes|0\rangle_{A}\otimes|0\rangle_{S}+ \sin \frac{\theta}{2}\otimes|0\rangle\otimes|0\rangle_{A} \otimes|1\rangle_{S}+|0\rangle\otimes|1\rangle_{A}\otimes|0\rangle_{S})  \\
&& \equiv  \frac{1}{\sqrt{2}}\left[\cos \frac{\theta}{2}|0\rangle\otimes|0\rangle_{A}\otimes|0\rangle_{S}+|0\rangle\otimes \left( \sin \frac{\theta}{2}|0\rangle_{A}\otimes|1\rangle_{S}+|1\rangle_{A}\otimes|0\rangle_{S}\right)\right] \\
\end{eqnarray*}

After the third event, the time-dependent field amplitudes $\hat a^f_-(t)$ and $\hat a^f_+(t)$ of the emitted radiation in waveguides~A and~S, respectively, are read-out simultaneously over a span of 1.02~$\mu$s. From here onwards, we follow the experimental and computational scheme of Ref.~\cite{eichler2011a} for all subsequent signal processing and analysis. The measured signals are digitally processed in real-time by integrating them over weighted time windows based on the temporal profile of the emitted photons of respective waveguide, a procedure known as temporal mode matching. The result is single-shot measurements of the time-\textit{independent} photonic modes $\hat{a}_-$ and $\hat{a}_+$, superimposed over background noise modes, represented as $\hat{h}_{-,+}^\dagger$. The noise modes are measured separately by reading-out in the absence of the signal when no pulse is sent to the system and then eliminated away computationally. Because of our hardware limitation of 1.02~$\mu$s in acquisition period, we capture the available signal with only an efficiency of about 74.8\% and 98.9\% for the waveguides~A and~S, respectively (see Fig.~S4). The acquisition efficiency and the total gain provided in the measurement chain to the signal was accounted for by a normalization coefficient determined from the obtained data in the main text. We have also accounted for the relatively high energy relaxation rates owing to the large direct coupling rates in the normalization, with the help of Linblad-based master equation and input-output theory. Based on reflection spectroscopy of the two modes (Fig.~2) and their analyses above, we realize all the fitting results of $\{\Gamma_i, \Gamma'_i\}$ for $i=\{s,a\}$ are completely accounted for by the two waveguides, bringing us to the conclusion that the pure dephasing rate is negligible. Therefore, the pure dephasing rate has been ignored in this normalization. We have captured 10 million single shots of each photonic mode to compute their statistical moments. The expected functional forms of all the statistical moments shown in the main text are (in the case of the $\pi/2$-pulse sent to $\ket{a}$):

\begin{align}
\langle \hat a_- \rangle & = \frac{1}{2} \cos \theta\\
\langle \hat a_+ \rangle & = \frac{1}{4} \sin \theta\\
\langle \hat a^\dagger_- \hat a_- \rangle & = \frac{1}{2}\\
\langle \hat a^\dagger_+ \hat a_+ \rangle & = \frac{1}{2} \sin^2 \frac{\theta}{2}\\
\langle \hat a^\dagger_- \hat a_+ \rangle & = \frac{1}{2} \sin \frac{\theta}{2} \\
\langle \hat a_- \hat a_+ \rangle & = 0
\end{align}

\subsection{Full Hamiltonian and transition dipole moments}

The full Hamiltonian of the two nominally-identical transmons coupled to each other is given by

\begin{equation}
H = \omega \hat{b}^\dagger_1 \hat{b}_1 + \omega \hat{b}^\dagger_2 \hat{b}_2 + \frac{\alpha}{2} \hat{b}^\dagger_1 \hat{b}^\dagger_1 \hat{b}_1 \hat{b}_1 + \frac{\alpha}{2} \hat{b}^\dagger_2 \hat{b}^\dagger_2 \hat{b}_2 \hat{b}_2 + g (\hat{b}^\dagger_1 \hat{b}_2 + \hat{b}^\dagger_2 \hat{b}_1)
\end{equation}

where $\hat{b}^\dagger_i$ and $\hat{b}_i$ are the creation and annihilation operators for transmon $i = \{1,2\}$; $\omega$ and $\alpha$ are the mode frequency and anharmonicity of each transmon; $g$ is the inter-transmon coupling rate~\cite{koch2007a}. The eigenstates and eigenvalues resulting from the diagonalization of the Hamiltonian up to two-excitations manifold are presented in Table~\ref{tbl:eigen}.

\begin{table}[h]
\centering
    \begin{tabular}{ l | l | c | c }
    \hline
    Eigenstate & Bare states composition & Eigenvalue & Value/$2\pi$ \\
    \hline
    $\ket{0}$ & $\ket{0,0}$ & $0$ & 0 GHz\\ 
    $\ket{a}$ & $\ket{1,0}-\ket{0,1}$ & $\omega - g$ & 5.6981 GHz\\
    $\ket{s}$ & $\ket{1,0}+\ket{0,1}$ & $\omega + g$ & 6.2909 GHz\\
    $\ket{2+}_L$ & $\ket{2,0}+\ket{0,2} - \frac{\alpha+\sqrt{16g^2+\alpha^2}}{2\sqrt{2}g}\ket{1,1}$ &$\frac{1}{2}(4\omega +\alpha -\sqrt{16g^2+\alpha^2})$ & 11.26 MHz\\
    $\ket{2-}$ & $\ket{2,0}-\ket{0,2}$ & $2\omega + \alpha$ & 11.7421 GHz\\
    $\ket{2+}_U$ & $\ket{2,0}+\ket{0,2} - \sqrt{2}\frac{\alpha-\sqrt{16g^2+\alpha^2}}{4g}\ket{1,1}$ &$\frac{1}{2}(4\omega +\alpha +\sqrt{16g^2+\alpha^2})$ & 12.4711 GHz\\
    \hline
    \end{tabular}
    \caption{Eigentates and eigenvalues of diagonalized Hamiltonian.}\label{tbl:eigen}
\end{table}

We found from spectroscopic data that $\omega/2\pi = 5.9945$~GHz, $\alpha/2\pi=246.9$~MHz and $g/2\pi = 296.4$~MHz, assuming identical transmons.

In the rotating frame, the classical field drives from the two waveguides~S and ~A are given by $b_S + b^\dagger_S$ and $b_A + b^\dagger_A$ respectively, in which $b_S = b_1 + b_2$ and $b_A = b_1 - b_2$. The transition dipole moments are, 

\begin{equation}
b_S + b^\dagger_S
= \begin{pmatrix}
0 & 0 & 2 & 0 & 0 & 0 \\
0 & 0 & 0 & 2 \sqrt{2} & 0 & 0 \\
2 & 0 & 0 & 0 & c_{S-} & c_{S+} \\
0 & 2 \sqrt{2} & 0 & 0 & 0 & 0 \\
0 & 0 & c_{S-} & 0 & 0 & 0 \\
0 & 0 & c_{S+} & 0 & 0 & 0
\end{pmatrix} = \begin{pmatrix}
0 & 0 & 2.00 & 0 & 0 & 0 \\
0 & 0 & 0 & 2.83 & 0 & 0 \\
2.00 & 0 & 0 & 0 & 0.53 & 6.31 \\
0 & 2.83 & 0 & 0 & 0 & 0 \\
0 & 0 & 0.53 & 0 & 0 & 0 \\
0 & 0 & 6.31 & 0 & 0 & 0
\end{pmatrix}
\end{equation}

\begin{equation}
b_A + b^\dagger_A
= \begin{pmatrix}
0 & 2 & 0 & 0 & 0 & 0 \\
2 & 0 & 0 & 0 & c_{A+} & c_{A_-} \\
0 & 0 & 0 & 2 \sqrt{2} & 0 & 0 \\
0 & 0 & 2 \sqrt{2} & 0 & 0 & 0 \\
0 & c_{A+} & 0 & 0 & 0 & 0 \\
0 & c_{A-} & 0 & 0 & 0 & 0 \\
\end{pmatrix} = \begin{pmatrix}
0 & 2.00 & 0 & 0 & 0 & 0 \\
2.00 & 0 & 0 & 0 & 5.13 & -0.65 \\
0 & 0 & 0 & 2.83 & 0 & 0 \\
0 & 0 & 2.83 & 0 & 0 & 0 \\
0 & 5.13 & 0 & 0 & 0 & 0 \\
0 & -0.65 & 0 & 0 & 0 & 0
\end{pmatrix}
\end{equation}

%
%

the second equality is obtained after evaluating the expressions,

\begin{align}
c_{S\pm}&=\frac{-\alpha \pm\sqrt{\alpha ^2+16 g^2}+4 g}{\sqrt{2} g} \\
c_{A\pm}&=\frac{\alpha \pm\sqrt{\alpha ^2+16 g^2}+4 g}{\sqrt{2} g}    
\end{align}

Fig.~S5 shows the transitions experimentally found with continuous-wave spectroscopy.

\begin{figure}
\includegraphics[width=1\linewidth]{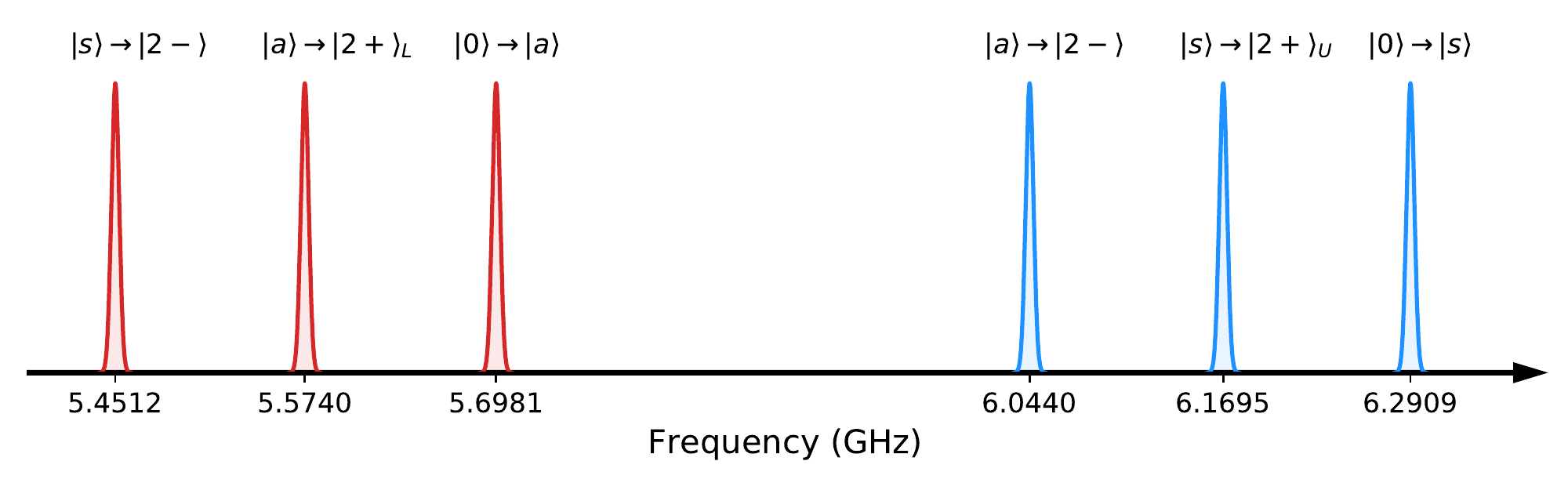}
\caption{All experimentally observed transition frequencies to scale. The blue (red) color indicates the transitions driven by waveguide~S~(A). 
}
\end{figure}

\subsection{Basic two-photon Raman model}

A basic Raman model for this system can be captured in the effective two-level basis of $\{|s\rangle, |a\rangle\}$ with an effective non-Hermitian Hamiltonian

\begin{equation}
H_R = 
\begin{pmatrix}
-2\frac{\Omega_+^2}{4\delta}-i (\Gamma_s + \Gamma'_s)/2 && -2\frac{\Omega_+\Omega_-}{4\delta} \\
-2\frac{\Omega_+\Omega_-}{4\delta} && -2\frac{\Omega_-^2}{4\delta}-i (\Gamma_a + \Gamma'_a)/2
\end{pmatrix}
\end{equation}

which takes into account all the coupling rates, $\{\Gamma_s, \Gamma'_s\}$, as decay rates of the states $\{|a\rangle, |s\rangle\}$. We calculate the reflectance and transmittance, when driving the state $\ket{s}$, from elements of the matrix $K(\omega)$, 

\begin{eqnarray}
& & K(\omega) = (H_R - \omega I)^{-1} \\
& & r(\omega) = \Gamma_s K_{11}(\omega) \\
& & T(\omega) = \sqrt{\Gamma_s \Gamma_a} K_{12}(\omega)
\end{eqnarray}

where I is the $2\times2$ identity matrix. Note that we use only the direct coupling rates $\Gamma_s$ and $\Gamma_a$ to the waveguide when calculating the $r$ and $t$. We find that $t(\omega)$ is maximized when the pump amplitudes, $\Omega_+ = \Omega_- = (\Gamma_a \Gamma_s)^{1/4} \sqrt{\delta}$ corresponding to a coherent population transfer.

\end{document}